%% file: main.tex
\documentclass[aps,prl,reprint,superscriptaddress,floatfix,nofootinbib]{revtex4-2}
\usepackage{verbatim}
\usepackage{mathtools}
\usepackage{ragged2e}
\usepackage{amsmath}
\usepackage{amsfonts}
\usepackage{amssymb}
\usepackage{graphicx}
\usepackage{slashed}
\usepackage{bm}
\usepackage{color}
\usepackage{epsf}
\usepackage{orcidlink}
\usepackage[parfill]{parskip}
\usepackage{nicefrac}
\usepackage{mathrsfs}
\usepackage{multirow}

\usepackage{hyperref}
\hypersetup{
  colorlinks   = true, 
  urlcolor     = blue, 
  linkcolor    = black, 
  citecolor   = red 
}

\newcommand{\M}{\mathcal{M}} 
\newcommand{\amp}{\mathcal{A}} 

\usepackage{tcolorbox}
\tcbuselibrary{listings, breakable, skins}
\newtcolorbox{eqbox}[2][]{%
	enhanced,
	colback=blue!5!white,
	colframe=blue!75!black,
	fonttitle=\bfseries,
	title=#2,
	sharp corners=south,
	boxrule=0.8pt,
	drop shadow,
	breakable,
	#1
}

\DeclareUnicodeCharacter{2212}{-}

\input{comments}

\begin{document}

\title{Quark and gluon tomography of the  helium-4 nucleus
}

\author{V.~Mart\'inez-Fern\'andez\,\orcidlink{0000-0002-0581-7154}}
\affiliation{Universit\'e Paris-Saclay - CEA - IRFU, 91191 Gif-sur-Yvette, France}
\affiliation{Center for Frontiers in Nuclear Science, Stony Brook University, Stony Brook, NY 11794, USA}
\author{B.~Pire\,\orcidlink{0000-0003-4882-7800}}
\affiliation{Centre de Physique Th\'eorique, CNRS, École polytechnique, I.P. Paris, 91128 Palaiseau, France  }

\author{P.~Sznajder\,\orcidlink{0000-0002-2684-803X}}
\affiliation{National Centre for Nuclear Research (NCBJ), Pasteura 7, 02-093 Warsaw, Poland}

\author{J.~Wagner\,\orcidlink{0000-0001-8335-7096}}
\affiliation{National Centre for Nuclear Research (NCBJ), Pasteura 7, 02-093 Warsaw, Poland}


\begin{abstract}
%
QCD collinear factorization allows coherent hard exclusive reactions to reveal the quark-gluon structure of light nuclei, enabling their 3D tomography. We study elastic form factors and deeply virtual Compton scattering on a helium-4 target, achieving theoretical precision unprecedented even in proton studies. Constraining generalized parton distributions at next-to-leading order in $\alpha_s$, incorporating kinematic twist corrections, and using full evolution equations, we provide the first tomography of a light nucleus, revealing distinct transverse spatial distributions of quarks and gluons.
\end{abstract}

\maketitle

\input{Introduction}

\input{CS}

\input{Model}

\input{Results}
\input{conclusions}
\input{acknowledgements}

\bibliography{bibliography}

\appendix


\end{document}

%% file: comments.tex
\newcounter{comment}

\definecolor{darkgreen}{rgb}{0,0.5,0}
\definecolor{orange}{rgb}{0.96,0.39,0}

%% file: Introduction.tex
{\bf Introduction.}
\label{sec::intro}
The coherent electroproduction of a photon off a nucleus $A$,
\begin{equation}\label{eq::electroproduction}
	e^-(k) + A(p)\to e^-(k') + A(p') + \gamma(q')\,,\quad
    q'^2 = 0\,,
\end{equation}
where the symbols in parentheses denote the four-momenta of the respective particles, is one of the most studied exclusive reactions used to access the quark and gluon structure of nucleons and nuclei. The importance of this process stems from the factorization theorem~\cite{Ji:1998pc}, which allows us to express the amplitude for the virtual Compton scattering,
\begin{equation}\label{eq::vcs}
	\gamma^*(q) + A(p)\to  \gamma(q') + A(p')\,,\quad
    q^2 < 0\,,\quad
    q'^2 = 0\,,
\end{equation}
where $q=k-k'$, in terms of generalized parton distributions (GPDs)~\cite{Diehl:2002he,Belitsky:2003fj}. GPDs encode correlations between the light-cone momentum sharing of the active parton and the four-momentum transfer to the target, provided that it remains intact. This enables access to the transverse position distributions of quarks and gluons within nucleons and nuclei, a technique referred to as hadron tomography~\cite{Berger:2001zb, Kirchner:2003wt, Guzey:2003jh,Cano:2003ju, Scopetta:2004kj}.
For the GPD framework to apply, one must consider scattering processes in the ``deep'' kinematic region. For reaction~\eqref{eq::electroproduction}, this means the virtuality $Q^2 = -q^2$ is large and scales with the squared center-of-mass energy $s = (q+p)^2$, while the squared four-momentum transfer $-t = -(p'-p)^2$ remains relatively small.

The helium-4 nucleus -- the alpha particle -- is a particularly clean and fundamental system for nuclear GPD studies. It is one of the lightest and most strongly bound nuclei, making it a natural testing ground for extending quark and gluon tomography from the nucleon to nuclei, where nuclear modifications of partonic distributions may become relevant. Its spin-0 and isospin-0 quantum numbers reduce the amplitude for the process~\eqref{eq::electroproduction} to a small set of GPDs. Furthermore, precise elastic form factor data~\cite{Repellin:1965vba,Frosch:1967pz,Arnold:1978qs,Ottermann:1985km,JeffersonLabHallA:2013cus} and existing CLAS measurements of coherent beam-spin asymmetries for this reaction~\cite{CLAS:2017udk} provide complementary constraints on the valence, sea-quark, and gluon sectors.  Consequently, helium-4 provides a unique target for the first quark
and gluon tomography of a light nucleus.

\begin{figure}
	\includegraphics[width=1\linewidth]{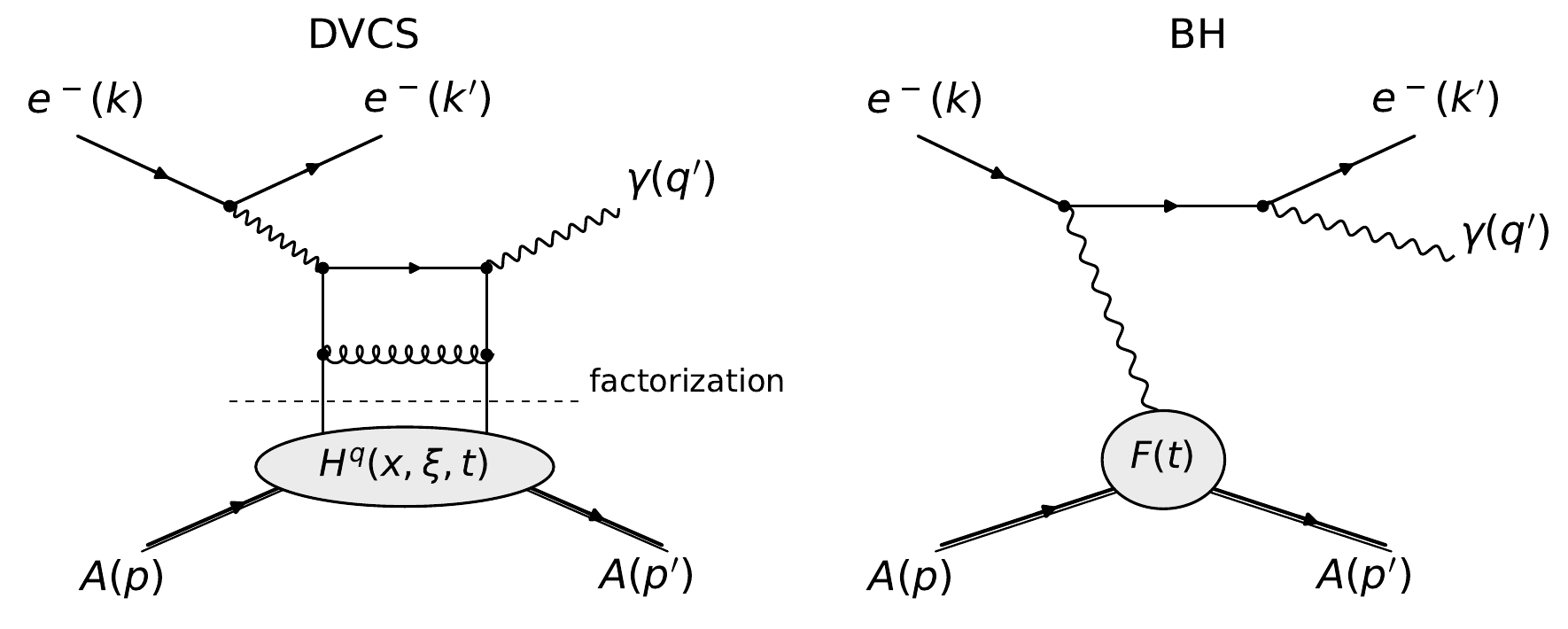}
	\caption{
Examples of Feynman diagrams contributing to exclusive photon electroproduction on a nuclear target: an NLO DVCS contribution (left panel) and the Bethe-Heitler (BH) contribution (right panel).}
    \label{fig::subprocesses}
\end{figure}

The scattering amplitude of the electroproduction reaction~\eqref{eq::electroproduction} is the sum of two contributions depicted in Fig.~\ref{fig::subprocesses}. The first one is the deeply virtual Compton scattering (DVCS) amplitude, which describes the QCD subprocess in which quarks and gluons actively participate in the interaction. This amplitude can be parameterized in terms of helicity amplitudes, given by convolution integrals of GPDs. The second contribution is the Bethe-Heitler (BH) amplitude, in which the final-state photon is emitted from the lepton line through bremsstrahlung. In contrast to the DVCS amplitude, the BH amplitude is purely electromagnetic and depends only on the elastic form factors of the target. 

GPDs are matrix elements of non-local light-cone quark-quark or gluon-gluon operators, defined separately for each quark flavor and for the gluons, as well as for specific spin configurations. For helium-4, which is the object of interest in our analysis, its spin-0 and isospin-0 nature means we will only consider the following unpolarized GPDs: $H^u(x,\xi,t) = H^d(x,\xi,t)$, $H^s(x,\xi,t)$, and $H^g(x,\xi,t)$; as well as one GPD representing the distribution of linearly polarized gluons: $H^g_T(x,\xi,t)$. Here, $x$ and $\xi$ denote the average light-cone momentum fraction carried by the active parton and the change in this fraction, respectively, while $t$ is the previously defined Mandelstam variable. GPDs also depend on the factorization scale, $\mu^2$, which will be omitted here and thereafter for brevity.

In the following sections, we first establish the link between GPDs and the DVCS cross section, while also commenting on the BH process. We then construct helium-4 GPD models and constrain them using elastic form-factor data and DVCS beam-spin asymmetries. The analysis is performed at an extended theoretical precision, including NLO corrections in $\alpha_S$, power-suppressed kinematical corrections, and full GPD evolution. We show that this accuracy is needed for a quantitative description of the available data, and we use the constrained GPDs to perform the first quark and gluon tomography of \mbox{helium-4}.

%% file: CS.tex
{\bf Cross section.} The scattering cross section for the process~\eqref{eq::electroproduction} takes the following form:
\begin{equation}\label{eq::cross_section}
    \frac{d^{4}\sigma^{s\lambda}_\chi}{d{x_A}\,dQ^2\,dt\,d\phi}
	= \frac{\alpha_{\rm em}^3}{8\pi} \frac{x_A y^2}{Q^4\sqrt{1+\omega^2}}
	\left|\frac{\mathcal{M}^{s\lambda}_\chi}{e^3}\right|^2 \,,
\end{equation}
where $\chi$ and $s$ denote the charge sign and helicity of the beam, respectively, while $\lambda$ denotes the polarization of the photon. In addition, $\alpha_{\rm em}$ is the fine-structure constant, $e$ the elementary charge, $y=(p \cdot q)/(p \cdot k)$ the inelasticity variable, $\omega = 2x_A M/Q$, $M$ the target mass, and the nuclear Bjorken variable $x_A= Q^2/(2p \cdot q)$.
The cross section also depends on a single angle, $\phi$, which in the general case is defined as the angle between the plane spanned by the leptons and the plane spanned by the nuclei~\cite{Bacchetta:2004jz}.

The amplitude appearing in Eq.~\eqref{eq::cross_section} is given by
\begin{align}\label{eq::total_iM}
	\M^{s\lambda}_{\chi}  
    =
    \M^{s\lambda}_{\chi,\,\rm DVCS}\left(\amp^{++}, \amp^{+-}, \amp^{0+}\right) 
    + \M^{s\lambda}_{\rm BH}(F) \,,
\end{align}
where the contributions from the DVCS and BH subprocesses are explicitly separated. While the BH amplitude is straightforward to evaluate and depends solely on the elastic form factors (with only a single form factor, $F(t)$, required for helium-4), the calculation of the DVCS amplitude is considerably more involved. 
Using the Kleiss-Stirling (KS) spinor techniques~\cite{KLEISS198461,KLEISS1985235}, we express it in terms of three helicity amplitudes: $\amp^{++}$, $\amp^{+-}$, and $\amp^{0+}$,
where $\amp^{\Lambda\Lambda'}$ describes the photon helicity transition $\Lambda\to\Lambda'$. Full expressions for $\M^{s\lambda}_{\chi,\,\rm DVCS}$ and $\M^{s\lambda}_{\rm BH}$ can be found in the accompanying paper~\cite{Martinez-Fernandez:2026web}.

The helicity amplitudes encode the GPD-dependent part of the process and are evaluated perturbatively through two distinct expansions. The first is the expansion in powers of the strong coupling constant $\alpha_S$, associated with the hard scattering kernel of the interaction, see Fig.~\ref{fig::subprocesses}. The second is the twist expansion in powers of $\sqrt{-t}/Q$. The latter contains two types of contributions: genuine higher-twist terms, involving new GPDs, and kinematic higher-twist terms, which depend only on leading-twist GPDs and reflect finite-kinematics effects, i.e. deviations from the Bjorken limit $Q^2\to\infty$ at fixed $x_A$. In the present study we restrict ourselves to the kinematic higher-twist sector, which is completely determined by leading-twist GPDs; genuine higher-twist contributions, involving additional nonperturbative functions, are left for future work.

In this work, we employ next-to-leading order (NLO) corrections in 
$\alpha_S$~\cite{Ji:1998xh} and kinematical twist-3 and twist-4 
corrections, suppressed by powers of $\sqrt{-t}/Q$ and 
$t/Q^2$~\cite{Braun:2012bg}. Previous phenomenological analyses 
have shown that NLO and kinematical higher-twist corrections can have 
a sizable impact on Compton observables, especially at fixed-target 
kinematics~\cite{Moutarde:2013qs,Martinez-Fernandez:2025gub}. At this accuracy, the helicity amplitudes take the following form:
\begin{align}
	\amp^{++}(\xi,t) = &
    \,\amp^{++}_{LO,LT}(H^\Sigma) 
    +
    \frac{\alpha_S}{4\pi}\amp^{++}_{NLO,LT}  (H^\Sigma, H^g)  \nonumber \\
    + & \frac{t}{Q^2}\amp^{++}_{LO,HT}(H^\Sigma) 
	\,, \label{eq::A++} \\
	\amp^{+-}(\xi,t) = &
    \,\frac{t}{Q^2}\amp^{+-}_{LO,HT}(H^\Sigma)
    +\frac{\alpha_S}{4\pi}\amp^{+-}_{NLO,LT}  (H_T^g)  
    \,, \label{eq::A+-} \\
	\amp^{0+}(\xi,t) = &
    \,\frac{\sqrt{-t}}{Q}\amp^{0+}_{LO,HT}(H^\Sigma)
    \,, \label{eq::A0+}
\end{align}
where
$H^\Sigma(x,\xi,t)= \sum_q e_q^2\,H^q (x,\xi,t)$ is the charge-weighted quark combination,
with $e_u = 2/3$ and  $e_d = e_s = -1/3$.
The explicit expressions for the building blocks entering the helicity amplitudes are given in the accompanying paper, Ref.~\cite{Martinez-Fernandez:2026web}. At leading order and leading twist, only the helicity-conserving amplitude $\amp^{++}$ is non-vanishing, and it depends solely on the quark GPDs  combination $H^\Sigma$. Gluon GPDs enter first at NLO in $\alpha_S$: the unpolarized gluon GPD $H^g$ contributes to $\amp^{++}$, while the gluon transversity GPD $H_T^g$ contributes to the photon-helicity-flip amplitude $\amp^{+-}$. The kinematical higher-twist corrections considered here generate additional power-suppressed contributions controlled by $t/Q^2$ and $\sqrt{-t}/Q$, including the longitudinal-transverse amplitude $\amp^{0+}$. Terms that simultaneously combine NLO corrections in $\alpha_S$ with kinematical higher-twist effects are beyond the accuracy adopted in this work and are therefore not included.

%% file: Model.tex
{\bf GPD model}.
From Eqs.~\eqref{eq::A++} to \eqref{eq::A0+}, we conclude that the cross-section~\eqref{eq::cross_section} is sensitive to the unpolarized GPDs $H$ for quarks and gluons, and to the transversity GPD $H_T$ for gluons only. Our model for helium-4 GPDs, $H^{i}$ (where $i = \{u_{\mathrm{val}}, u_{\mathrm{sea}}, d_{\mathrm{val}}, d_{\mathrm{sea}}, s, g\}$ denotes the parton type, explicitly distinguishing between valence and sea components), is based on double distributions (DDs) $F^{i}$.
The GPDs are related to these DDs via a Radon transform~\cite{Radyushkin:1998es}:
\begin{equation}
H^{i}(x,\xi,t) =
\int_{-1}^{1}d\beta
\int_{-1+|\beta|}^{1-|\beta|}d\alpha \,
\delta(\beta + \xi\alpha - x)
F^{i}(\beta, \alpha, t) \,,
\end{equation}
where we have neglected the so-called $D$-term~\cite{Polyakov:1999gs} as it is not relevant for the data analyzed in this work. Modeling based on DDs helps to fulfill polynomiality, a fundamental property 
of GPDs that follows from Lorentz covariance and is otherwise non-trivial to implement.

The construction of DDs is as follows:
\begin{equation}
F^{i}(\beta, \alpha, t) = f^A_{i}(\beta, t)\,h_{i}(\beta, \alpha) \,,
\label{eq::dd_master}
\end{equation}
where $f^A_{i}(\beta, t)$ is a $t$-dependent generalization of parton distribution function (PDF) and $h_{i}(\beta, \alpha)$ is the profile function,
\begin{equation}
h_{i}(\beta, \alpha) = 
\frac{\Gamma(2b_{i}+2)}{2^{2b_{i}+1}\Gamma^2(b_{i}+1)}
\frac{\left( (1-|\beta|)^2 - \alpha^2\right)^{b_{i}}}{(1-|\beta|)^{2b_{i}+1}}\,,
\label{eq:dd_profile}
\end{equation}
with $b_{i} = 1$ for valence quarks and $b_{i} = 2$ for sea quarks and gluons. The profile function is responsible for generating the $\xi$-dependence of the final model, and its normalization ensures that $H^i(x,0,t) = f^A_i(x,t)$.

The $t$-dependent PDF is
\begin{equation}
f^A_{i}(\beta, t) = f^A_{i}(\beta)\, \frac{k_{i}(|\beta|, t)}{k_{i}(|\beta|, 0)} \,,
\end{equation}
where $f^A_{i}(\beta)$ accounts for both partons ($\beta > 0$) and antipartons ($\beta < 0$). For $A|\beta| \le 1$ we have
\begin{align}
f^A_{i}(\beta) = A^2
\begin{cases}
\Theta(\beta)\,f^{p/A}_{i}(A|\beta|) \,,
& \mathrm{for}~i = u_{\mathrm{val}}, d_{\mathrm{val}} \,, \\
\mathrm{sgn}(\beta)\,f^{p/A}_{i}(A|\beta|) \,, & \mathrm{for}~i = u_{\mathrm{sea}}, d_{\mathrm{sea}}, s \,, \\
|\beta|\,f^{p/A}_{i}(A|\beta|) \,, & \mathrm{for}~i = g \,,
\end{cases}
\label{eq::forwardFromPheno}
\end{align}
where $\Theta(\cdot)$ and $\mathrm{sgn}(\cdot)$ are the step and signum functions, respectively, while $f^{p/A}_{i}(A|\beta|)$ are the PDFs taken from the nNNPDF30 parameterizations~\cite{AbdulKhalek:2022fyi}. Since these parameterizations are defined with respect to the proton momentum and are normalized per nucleon, we rescale $|\beta|$ by $A$ (since $x_A \approx x_B/A$) and include an overall prefactor of $A$. To avoid probing the PDFs in unknown domains, we explicitly set $f^A_{i}(\beta)=0$ for $\beta>1/A$.

\begin{figure}[!ht] 
  \centering
  \includegraphics[width=0.4\textwidth]{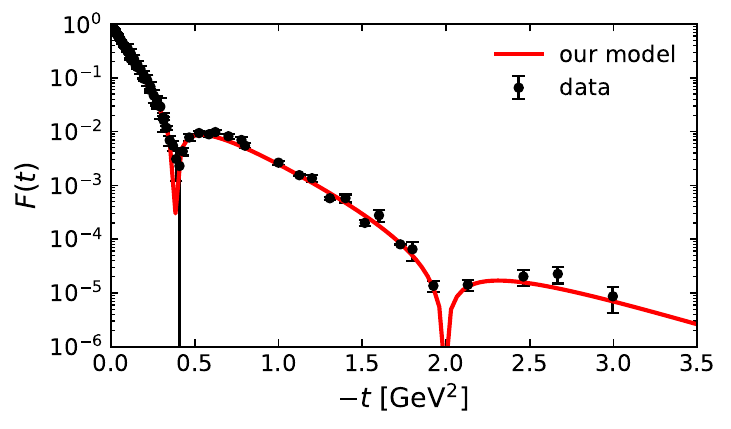}
  \caption{Fit to helium-4 elastic form factor data.}
  \label{fig::eff_fit}
\end{figure}

We now focus on $k_{i}(|\beta|, t)$, which is responsible for the dependence on the variable $t$, and is crucial for the hadron tomography. For valence quarks we use the following novel Ansatz
\begin{align}
k_{i}(|\beta|, t)& =  
\left(\frac{1}{1-p_0(1-|\beta|)^2t}\right)^{p_1} \nonumber\\
& \times \prod_{j=1}^{n}\left(\frac{|p_{2,j} + t|}{|p_{2,j} + t| - p_{3,j}(1-|\beta|)^2t}\right)^{p_{4,j}}\,.
\label{eq::tDep}
\end{align}
The free parameters $p_j$ are fixed in a fit to elastic form factor data~\cite{Repellin:1965vba,Frosch:1967pz,Arnold:1978qs,Ottermann:1985km,JeffersonLabHallA:2013cus}, utilizing the following relation:
\begin{equation}
F(t) = \frac{1}{Z} \int_{-1}^{1}dx \left(e_{u}f_{u_{\mathrm{val}}}^A(x, t) + e_{d}f_{d_{\mathrm{val}}}^A(x, t)\right)\,,
\label{eq::eff}
\end{equation}
where $Z$ is the atomic number of helium-4. We capture the first two diffractive minima, $n=2$. The result of our fit is shown in Fig.~\ref{fig::eff_fit}.

\begin{figure}[!ht] 
  \centering
  \includegraphics[width=0.4\textwidth]{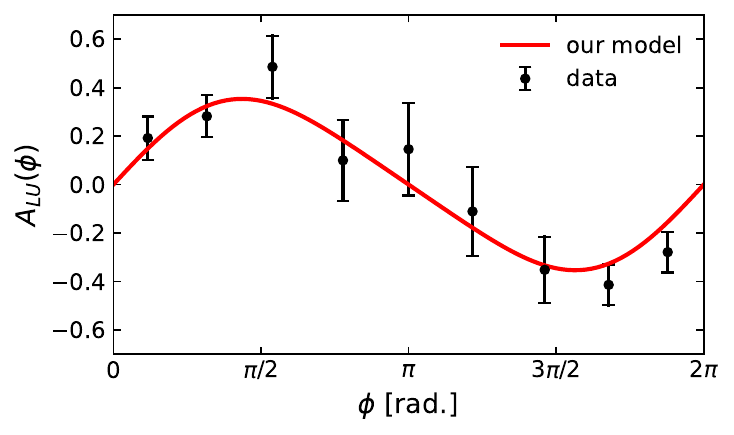}
  \caption{Fit to the beam spin asymmetry for DVCS, $A_{LU}$, measured on a helium-4 target by CLAS~\cite{CLAS:2017udk}. Only one kinematic bin is shown here: $x_B=0.172$, $t=-0.099\,\mathrm{GeV}^2$ and $Q^2=1.42 \,\mathrm{GeV}^2$.}
  \label{fig::dvcs_fit}
\end{figure}

For sea quarks and gluons, on the other hand, we adopt a simple exponential form:
\begin{equation}
k_{i}(|\beta|, t) = \exp\big(p(1-|\beta|^2)t\big) \,,
\label{eq::tDepSea}
\end{equation}
containing only a single parameter $p$. We make the standard assumption that the spatial transverse profiles of sea quarks and gluons identical. Challenging this assumption would require a multichannel analysis of exclusive scattering off helium-4, while adopting a more complex ansatz than Eq.~\eqref{eq::tDepSea} demands more DVCS data probing lower values of $x_A$ and $t$ than are currently available. The parameter $p$ is determined by fitting to JLab DVCS data for the beam-spin asymmetry~\cite{CLAS:2017udk}, as illustrated for a single kinematic bin in Fig.~\ref{fig::dvcs_fit}. Our nominal fit is performed at NLO/HT precision, calculating the amplitudes via Eqs.~\eqref{eq::A++}-\eqref{eq::A0+}. The elastic form factor required to evaluate the BH contribution is calculated directly from the GPDs using Eq.~\eqref{eq::eff}. We employ the full evolution equations for GPDs~\cite{Bertone:2017gds}, allowing us to accurately match our predictions to the various $Q^2$ kinematics of the measured data. The impact of the extended theoretical accuracy is quantified in Table I.
Fits performed at lower precision, LO/LT or NLO/LT, lead to a poorer
description of the data, whereas the NLO/HT setup reaches
$\chi^2/N_{\text{pts}}\simeq 1$.
\begin{table}[!htbp]
\caption{Comparison of our $A_{LU}$ fits to CLAS data~\cite{CLAS:2017udk} at increasing theoretical precision in the $\alpha_s$ and kinematical-twist expansions. The listed GPDs and helicity amplitudes specify the ingredients of each setup, while the last column gives $\chi^2/N_{\text{pts}}$.}
\label{tab:summary}
\begin{ruledtabular}
\begin{tabular}{cccc}
Precision & GPDs & Amplitudes & $\chi^2/N_{\text{pts}}$ \\
\colrule
LO/LT & $H^q$ 
    & $\amp^{++}$ 
    & 1.31 \\

NLO/LT & $H^q$, $H^g$, $H_T^g$ 
    & $\amp^{++}$, $\amp^{+-}$ 
    & 1.28 \\

NLO/HT & $H^q$, $H^g$, $H_T^g$ 
    & $\amp^{++}$, $\amp^{+-}$, $\amp^{0+}$ 
    & 1.00 \\
\end{tabular}
\end{ruledtabular}
\end{table}

Finally, we address the modeling of the GPD $H_T^g$. Since little is known about this GPD, we approximate it by the unpolarized GPD $H^g(x,\xi,0)$ as $t \to 0$, while for $t \ll 0$ we use the positivity inequality~\cite{Pire:1998nw, Kirch:2005in}. This results in a conservative model,
\begin{equation}
   H_{T}^{g}(x, \xi, t) = H^{g}(x, \xi, 0)\sqrt{\frac{1}{1 - t/(4M^2)}}\,,
\end{equation}
where at $t=0$ one has $H_{T}^{g}(x, \xi, 0) = H^{g}(x, \xi, 0)$, whereas for $|t| > 0$ the model exhibits a much milder dependence on $t$ than $H^{g}(x, \xi, t)$. Under these conditions, we find that the contribution of $H_{T}^{g}(x, \xi, t)$ to the $\amp^{+-}$ helicity flip amplitude is negligible compared to the kinematical higher twist contribution.

%% file: Results.tex
{\bf Results.}
Hadron tomography reveals the spatial distribution of partons carrying a specific fraction of the hadron's longitudinal momentum in the transverse plane. It is obtained by Fourier-transforming the unpolarized GPDs at $\xi=0$ with respect to the transverse momentum transfer ($t = -\boldsymbol{\Delta_T}^2$):
\begin{equation}
f^A_i(x,\boldsymbol{b_T}) =\int \frac{\mathrm{d}^2\boldsymbol{\Delta_T}}{(2\pi)^2}e^{-i\boldsymbol{b_T}\cdot \boldsymbol{\Delta_T}}H^i(x,0,-\boldsymbol{\Delta_T}^2) \,.
\end{equation}
Here, $\boldsymbol{b_T}=(b_x, b_y)$ is the transverse impact parameter, defined in a coordinate system whose origin coincides with the center of momentum of the hadron's constituents~\cite{Burkardt:2002hr}. 

\begin{figure}[!ht] 
  \centering
  \includegraphics[width=0.5\textwidth]{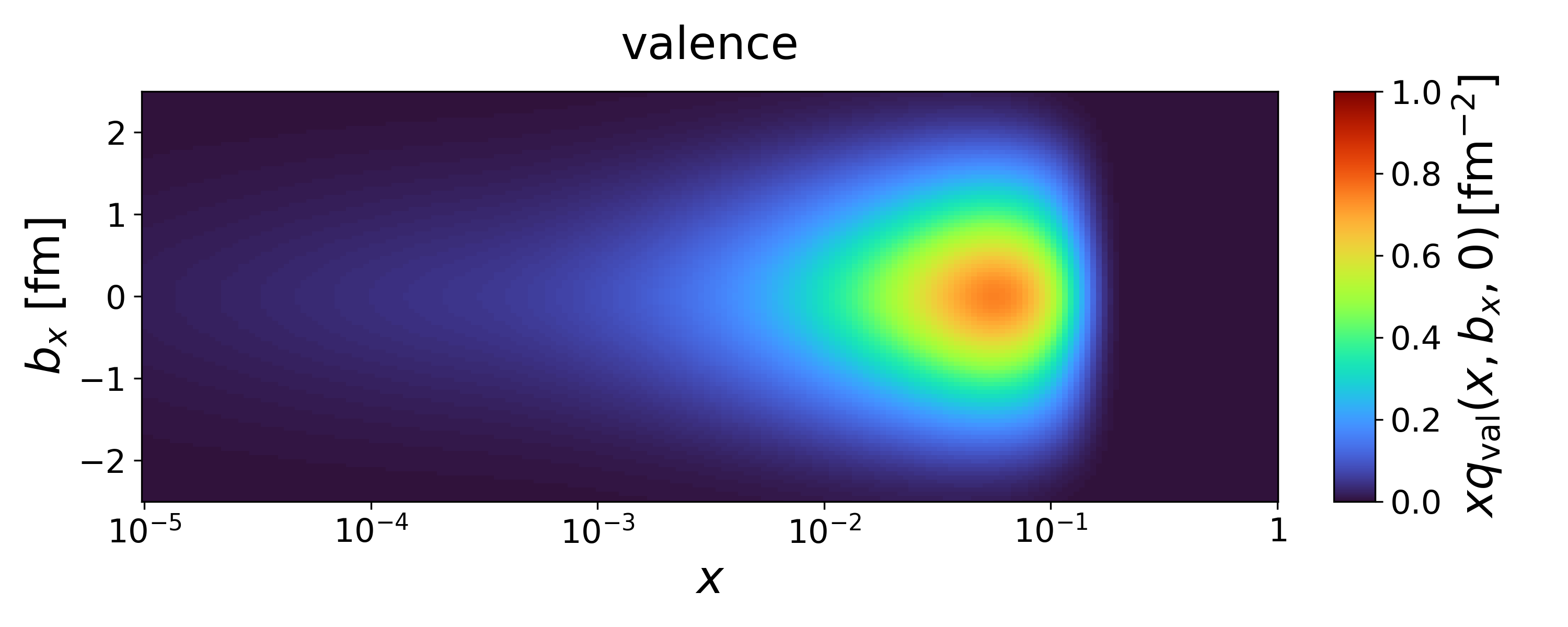}
  \includegraphics[width=0.5\textwidth]{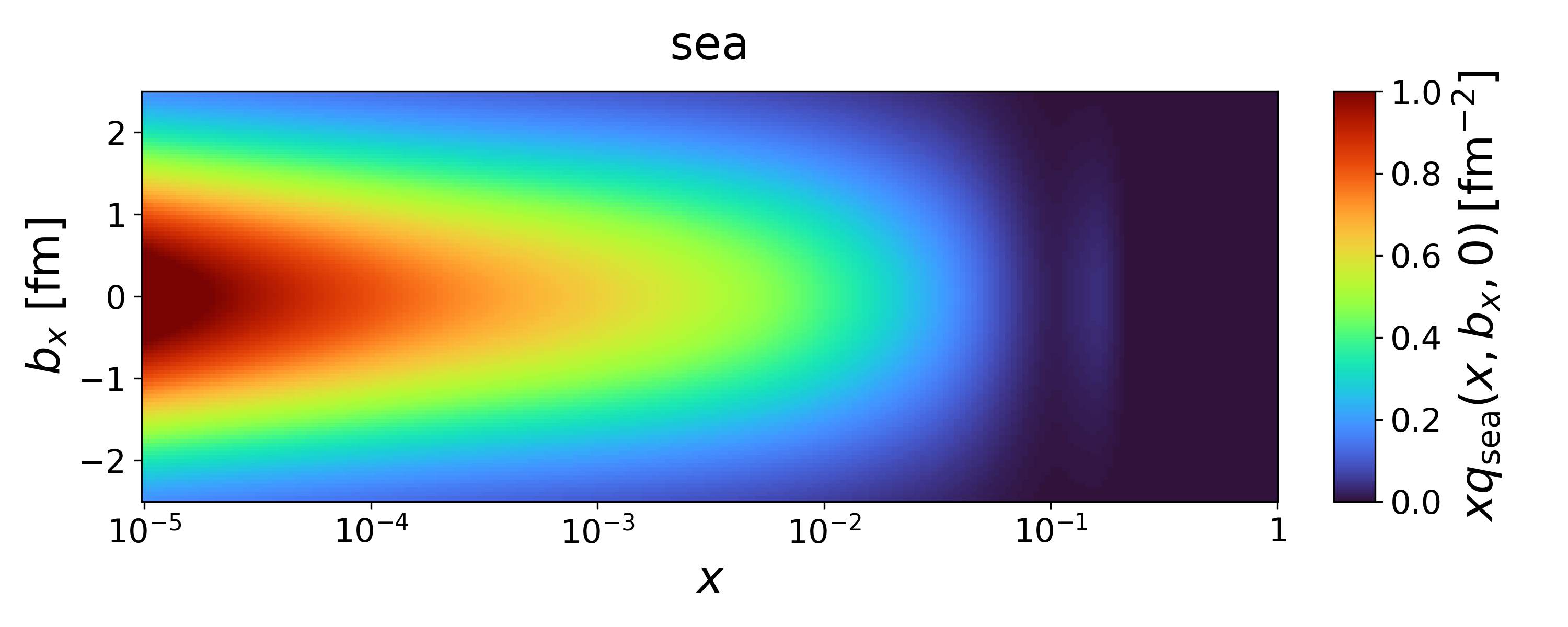}
  \includegraphics[width=0.5\textwidth]{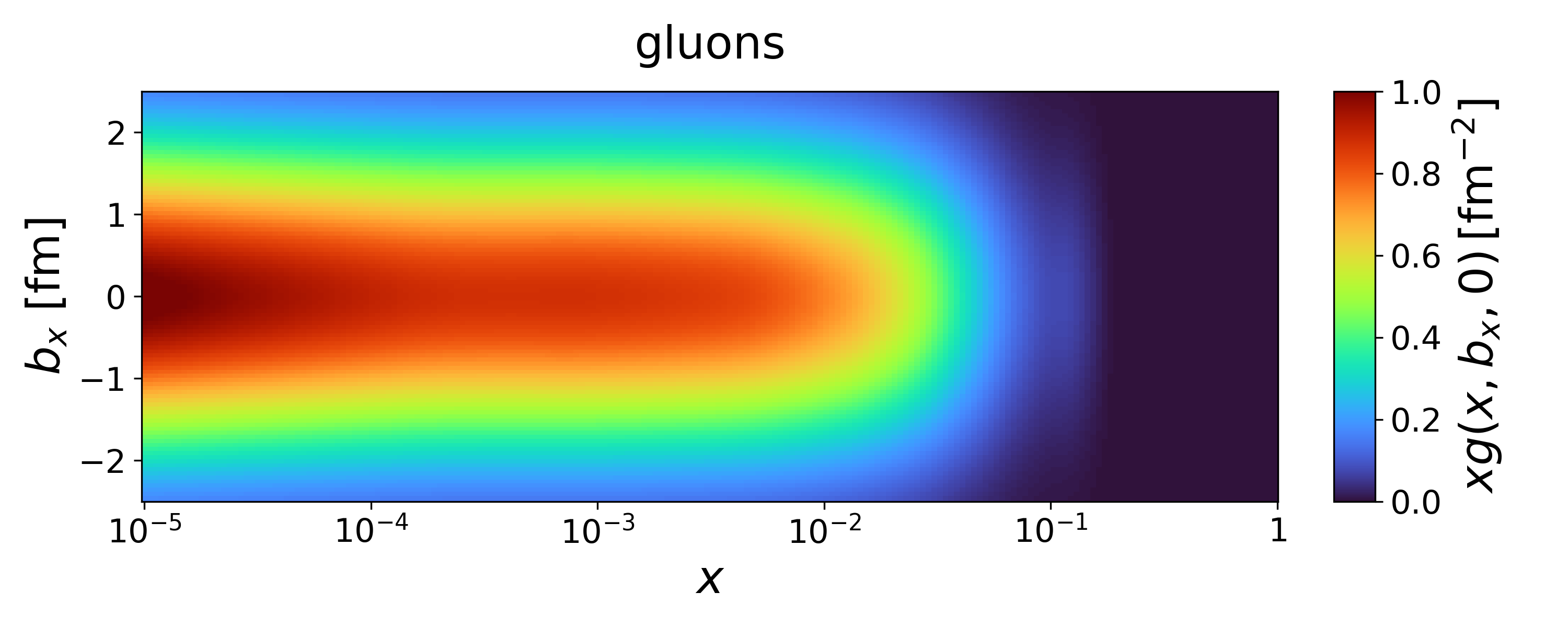}
  \caption{Spatial distributions of quarks, $x\,q(x, b_x, b_y=0)$ (separately for valence and sea contributions), and gluons, $x\,g(x, b_x, b_y=0)$, in helium-4 nuclei at the scale $\mu^2 = 2\,\mathrm{GeV}^2$.}
  \label{fig::nt}
\end{figure}

\begin{figure}[!ht] 
  \centering
  \includegraphics[width=0.4\textwidth]{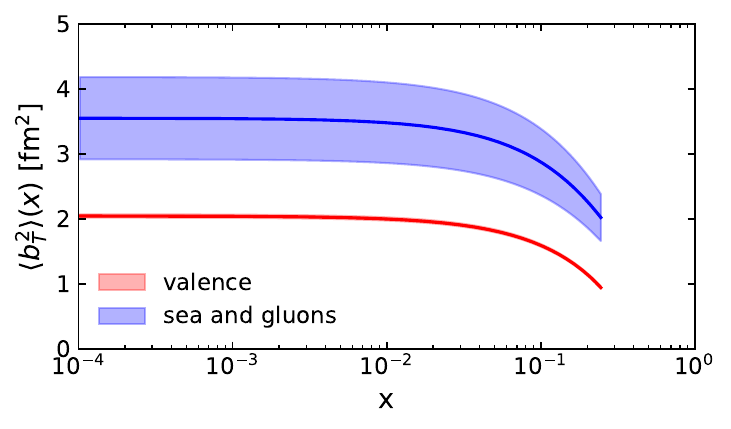}
  \caption{Normalized second moments of spatial distributions (see Eq.~\eqref{eq::r2_x}) for quarks and gluons in helium-4 nuclei at the scale $\mu^2 = 2\,\mathrm{GeV}^2$.}
  \label{fig::nt_r2_x}
\end{figure}

The tomographic profiles obtained using our constrained GPD model are shown in Fig.~\ref{fig::nt}, separated into contributions from valence quarks, sea quarks, and gluons.
At a given $x$, the transverse extension of partons can be quantified by the second normalized moment of the transverse distribution,
\begin{equation}
    \langle b_T^2 \rangle_i(x) = \frac{\int d^2 \boldsymbol{b_T} \boldsymbol{b_T}^2 f^A_i(x, \boldsymbol{b_T})}{\int d^2 \boldsymbol{b_T} f^A_i(x, \boldsymbol{b_T})} \,.
    \label{eq::r2_x}
\end{equation}
This quantity is shown in Fig.~\ref{fig::nt_r2_x}, including uncertainties obtained by mixing replicas of PDF, elastic, and DVCS data. The result is identical for sea quarks and gluons, which is a consequence of using the same profile function~\eqref{eq::tDepSea} in both cases.

The extracted transverse profiles reveal a clear hierarchy. The spatial distribution of sea quarks and gluons is broader than that of valence quarks, a trend also observed in the proton (see, e.g., Ref.~\cite{Goloskokov:2006hr}). Because the employed nuclear PDFs are defined only in the region $x\leq 1/A$,
the tomographic plots are not populated above $x=1/4$. Furthermore, we observe that the spatial distribution becomes narrower as $x$ grows. This effect is an inherent consequence of the reference system in which the tomography is defined: in the $x \to 1$ limit, the entire momentum is carried by a single parton located at the origin of the coordinate system. Our ansatz correctly reproduces this theoretical constraint. Finally, the overall distribution is naturally wider than that of the proton. 

The overall radii can be quantified by
\begin{align}
    \langle b_T^2 \rangle_q = \frac{\int d x \int d^2 \boldsymbol{b_T} \boldsymbol{b_T}^2 f^A_q(x, \boldsymbol{b_T})}{\int d x \int d^2 \boldsymbol{b_T} f^A_q(x, \boldsymbol{b_T})} \,, \\
    \langle b_T^2 \rangle_g = \frac{\int \frac{d x}{x} \int d^2 \boldsymbol{b_T} \boldsymbol{b_T}^2 f^A_g(x, \boldsymbol{b_T})}{\int \frac{d x}{x} \int d^2 \boldsymbol{b_T} f^A_g(x, \boldsymbol{b_T})} \,.
    \label{eq::r2}
\end{align}
Due to isospin symmetry, for helium-4 the quantities $\langle b_T^2 \rangle_{q_{\mathrm{val}}}(x)$ and $\langle b_T^2 \rangle_{q_{\mathrm{val}}}$ yield the same values regardless of whether they are evaluated using the unweighted combination, $f^A_{u_{\mathrm{val}}}(x,\boldsymbol{b_T}) + f^A_{d_{\mathrm{val}}}(x,\boldsymbol{b_T})$, or the charge-weighted combination, $e_u\,f^A_{u_{\mathrm{val}}}(x,\boldsymbol{b_T}) + e_d\,f^A_{d_{\mathrm{val}}}(x,\boldsymbol{b_T})$. We obtain $\langle b_T^2 \rangle_{q_{\mathrm{val}}} = 1.943(28)\,\mathrm{fm}^2$ for valence quarks, $\langle b_T^2 \rangle_{q_{\mathrm{sea}}} = 3.51(63)\,\mathrm{fm}^2$ for sea quarks, and $\langle b_T^2 \rangle_{g} = 3.50(62)\,\mathrm{fm}^2$ for gluons. The value for valence quarks can be compared to the mean electric charge radius of helium-4 nuclei, $r^e_{{}^4\mathrm{He}}$, extracted from elastic form factor analyses. This comparison must account for the fact that the charge radius integrates over three spatial dimensions, whereas our evaluation is restricted to the transverse plane:
\begin{equation}
    r^e_{{}^4\mathrm{He}} = \sqrt{\frac{3}{2}\langle b_T^2 \rangle_{q_{\mathrm{val}}}}\,,
\end{equation}
which holds under the natural assumption of spherical symmetry.
We obtain $1.71(21)\,\mathrm{fm}$, which can be compared to the value of $1.6785(21)\,\mathrm{fm}$ reported in Ref.~\cite{Mohr:2024kco}. The larger uncertainty reflects a different approach, in which the radius is evaluated by analyzing the partonic content of the helium-4 nucleus. For context, the corresponding value for the proton is $r^e_{p} = 0.840 75(64)\,\mathrm{fm}$~\cite{Mohr:2024kco}.

%% file: conclusions.tex
{\bf Conclusions.
}
Using elastic form-factor and coherent DVCS data, we have performed
the first quark and gluon tomographic study of the helium-4 nucleus. The
analysis combines NLO corrections in $\alpha_S$, kinematic twist-3 and twist-4
corrections, and full GPD evolution, which are required for a quantitative
description of the existing fixed-target data. The resulting images reveal a
broader transverse distribution of sea quarks and gluons than of valence
quarks. We demonstrate that GPD-based analyses provide unique insights into the partonic structure of the target compared to traditional elastic form factors. 

In the near future, data will be collected at slightly higher energies with the $12\,\mathrm{GeV}$ electron beam at JLab~\cite{CLAS:2021ovm}, which should allow for more precise data over a broader $Q^2$ and $t$ range. This will constitute a stringent test of the validity of our approach. In the longer term, even more data should be available at  Electron-Ion Colliders (EIC and EIcC)~\cite{AbdulKhalek:2021gbh,Anderle:2021wcy}, allowing for better constraints on the intermediate and low $x_A$ kinematic domains.

{\bf Acknowledgements.}
We acknowledge useful discussions and correspondence with R.~Dupr\'e, V.~Guzey, and C.~Mezrag.
This research was funded in whole or in part by the National Science Centre, Poland (grant
IMPRESS-U No.~2024/06/Y/ST2/00155 and grant SONATA BIS-15 No.~2025/58/E/ST2/00045). For the purpose of open access, the authors have applied a CC-BY copyright licence to any Author Accepted Manuscript (AAM) version arising from this submission. The research of V.M.F. was funded in part by l’Agence Nationale de la Recherche (ANR), project ANR-23-CE31-0019.
